\documentclass[aps,prd,twocolumn,groupedaddress,nofootinbib]{revtex4-1}

\usepackage{dcolumn}

\usepackage{amssymb}

\usepackage{amsmath}

\usepackage{bm}

\newcommand{\beq}{\begin{equation}}
\newcommand{\eeq}{\end{equation}}
\newcommand{\bea}{\begin{eqnarray}}
\newcommand{\eea}{\end{eqnarray}}
\newcommand{\ben}{\begin{enumerate}}
\newcommand{\een}{\end{enumerate}}

\newcommand{\pa}{\partial}

\newcommand{\na}{\nabla}
\newcommand{\ed}{{\rm d}}
\newcommand{\ced}{{\rm D}}
\newcommand{\we}{\wedge}
\newcommand{\Lie}{{\cal L}}

\newcommand{\Tr}{{\rm Tr}}
\newcommand{\ti}{\tilde}
\newcommand{\Rs}{\mathbb{R}}
\newcommand{\Cs}{\mathbb{C}}

\renewcommand\({\left(}
\renewcommand\){\right)}
\renewcommand\[{\left[}
\renewcommand\]{\right]}

\newcommand{\os}{\overset}

\newcommand{\nn}{\nonumber}

\newcommand{\al}{\alpha}
\newcommand{\be}{\beta}
\newcommand{\ga}{\gamma}
\newcommand{\Ga}{\Gamma}
\newcommand{\de}{\delta}
\newcommand{\De}{\Delta}
\newcommand{\ep}{\epsilon}
\newcommand{\vep}{\varepsilon}

\newcommand{\et}{\eta}
\newcommand{\te}{\theta}
\newcommand{\Te}{\Theta}

\newcommand{\La}{\Lambda}
\newcommand{\ro}{\rho}
\newcommand{\si}{\sigma}
\newcommand{\Si}{\Sigma}
\newcommand{\ta}{\tau}
\newcommand{\ph}{\phi}

\newcommand{\om}{\omega}
\newcommand{\Om}{\Omega}

\begin{document}

\title{Differential form description of the Noether-Lagrange machinery, vielbein/gauge-field analogies and energy-momentum complexes}

\author{Ermis Mitsou}
\email{ermis.mitsou@unige.ch}
\affiliation{D\'epartement de Physique Th\'eorique and Center for Astroparticle Physics, Universit\'e de 
	     Gen\`eve, 24 quai Ansermet, CH--1211 Geneva, Switzerland}

\begin{abstract}

We derive the variational principle and Noether's theorem in generally covariant field theory in an explicitly coordinate-independent way by means of the exterior calculus over the space-time manifold. We then focus on the symmetry of active diffeomorphisms, that is, the pushforwards along the integral lines of any vector field, and its analogies with internal gauge symmetries. For instance, it is well known that a class of Noether currents associated to a gauge symmetry can be obtained by taking the {\it partial} derivative of the Lagrangian with respect to the corresponding gauge field. Here we show that this relation also holds for the Noether currents associated to diffeomorphisms and the vielbein, but only if one decomposes all forms in the vielbein basis. We also relate the diffeomorphism Noether currents to the matter energy-momentum tensor of General Relativity, to Hamiltonian boundary terms and to two known energy-momentum complexes of the vielbein.

\vspace{1cm}

\end{abstract}

%\pacs{03.50.-z, 04.20.-q, 04.20.Fy}
%\keywords{Noether's theorem, general covariance, energy-momentum complex}

\maketitle

\section{Introduction}

The aim of this paper is three-fold. First, we propose to review the basic Noether-Lagrange machinery of generally covariant field theory, i.e. the variational principle and Noether's theorem, both in terms of currents and charges, without ever referring to coordinates. We therefore employ the framework of exterior calculus over the space-time manifold $\cal M$, an explicitly coordinate-independent and global way of displaying and manipulating the field information\footnote{From the fibre bundle point of view of field theory, we will be using local coordinates for the vertical space, our interest being mainly to get rid of the space-time coordinates.}. Apart from elegance, sticking to this formalism guarantees that general covariance is always preserved and prevents us from being distracted by coordinate-induced artefacts. Another advantage is that it allows one to ``see" conservation equations just by looking at the Euler-Lagrange equations.  

Using some simple identities listed in appendix \ref{sec:useide}, we will see that many computations in General Relativity (GR) that are usually lengthy when carried out using coordinates become quite straightforward in exterior calculus. We illustrate this fact by computing the equations of motion and Noether currents of Palatini vielbein gravity coupled to a Yang-Mills (YM) and a Dirac field. 

This brings us to the second aim of our paper which is to draw some interesting analogies between vielbein gravity and classical gauge theory. For instance, it is well known that some Noether currents of YM theory can be computed by simply taking the partial derivative of the Lagrangian with respect to the gauge field. We will show that, {\it if} one decomposes all forms in the vielbein basis, then, in full analogy, one obtains the diffeomorphism Noether currents by taking the partial derivative of the Lagrangian with respect to the vielbein.

Finally, we focus on the concept of energy and momentum. We relate the Noether diffeomorphism currents of the matter Lagrangian to the energy-momentum tensor as defined in GR, that is, the variational derivative of the matter action with respect to the vielbein. We also discuss the relation between the Noether energy charge and the Hamiltonian definition of energy, i.e. boundary terms in the canonical formalism, and show that a Noether energy does not always have a Hamiltonian analogue. We conclude by discussing the relation of the Noether diffeomorphism currents and some known energy-momentum complexes for the gravitational field. For instance, in the case where one considers M\o ller's Lagrangian \cite{Mol} for GR, we show that the Noether currents corresponding to the diffeomorphisms generated by the vielbein and holonomic frames are nothing but the energy-momentum complexes of \cite{DVM, Sza} and the one of M\o ller \cite{Mol}, respectively. The former is a tensor under diffeomorphisms, but transforms inhomogeneously under local Lorentz transformations (LLTs), while the latter is non-covariant under both transformations. 

Our study also contains some useful by-products, such as an elegant expression of M\o ller's Lagrangian in terms of differential forms (\ref{eq:LM}), as well as a coordinate-independent and compact expression for the variation of the Hodge scalar product with respect to the vielbein (\ref{eq:starvar}).

The organization of the paper goes as follows. In section \ref{sec:setup} we introduce the notational conventions, symmetries and fields we are going to use. In section \ref{sec:basics} we derive the variational principle and Noether's theorem using only forms, anti-derivations and integration. In section \ref{sec:diffNo} we focus on the Noether currents associated to diffeomorphisms and their analogies with the case of internal gauge symmetries. In section \ref{sec:EM} we discuss the relation the matter energy-momentum tensor, the Hamiltonian boundary terms and some gravitational energy-momentum complexes. In section \ref{sec:conclusion} we summarize.

\section{Conventions and notation} \label{sec:setup}

\subsection{Geometry and groups}

Let $\cal M$ be a real parallelizable smooth manifold of dimension $D \equiv d + 1 \geq 4$ and let $\De^k$ denote the set of $k$-dimensional orientable closed embedded submanifolds of $\cal M$. For $U \in \De^D$, $\frak{X}(U) \equiv \Ga(TU)$ denotes the space of vector fields and $\Om^p(U) \equiv \Ga\( \La^p(U) \)$ denotes the space of $p$-forms, over $U$. If the $U$ dependence is omitted then it means the assertion works for all $U \in \De^D$. 

The parallelizability of $\cal M$ implies the existence of global frames, i.e. sets of $D$ vector fields $\ep_I \in \frak{X}(\cal M)$, where $I = 1, \dots, D$, which form a basis everywhere on $\cal M$. Given such a frame, there exists a unique set of $D$ vielbein $1$-forms $e^I$ obeying $e^I (\ep_J) = \de^I_J$ and they are non-singular $e^{I_1} \we \dots \we e^{I_D} \neq 0$ everywhere on $\cal M$. The Lorentzian metric field is defined by $g \equiv \et_{IJ}\, e^I \otimes e^J$, where $\et_{IJ}$ is the Minkowski metric with a signature of mostly pluses. We define the compact notations $e^{I_1 \dots I_p} \equiv e^{I_1} \we \dots \we e^{I_p}$, $\ti{e}_{I_1 \dots I_p} \equiv \vep_{I_1 \dots I_D}e^{I_{p+1} \dots I_D} /(D-p)!$, $i_{I_1 \dots I_p} \equiv i_{I_1} \dots i_{I_p}$ and $\Lie_I \equiv \Lie_{\ep_I}$, where for the Levi-Civita symbol we use the convention $\vep_{012\dots D} = - \vep^{012\dots D} = 1$, $i_I$ denotes the interior product and $\Lie_I$ the Lie derivative with respect to $\ep_I$. Replace $I \to \xi$ for an arbitrary vector field $\xi$. The Hodge dual is defined by 
\beq \label{eq:Hodge}
\star\, \al \equiv \frac{1}{p!} \, \ti{e}^{I_1 \dots I_p} i_{I_p \dots I_1} \al \, , \hspace{1cm}  \al \in \Om^p \, .
\eeq
We use $\ed$ to denote the exterior derivative and $\ced$ for the covariant exterior derivative with respect to all internal gauge transformations. We let $\bar{\ed}$ denote the codifferential $\bar{\ed} \al \equiv (-1)^{D (p-1)} \star \ed \star \al$, where $\al \in \Om^p$. We adopt the ``anti-hermitian" convention for the Lie algebra of a Lie group, i.e. the latter is the exponentiation of the former with no $i$ factor. For the case of SU$(N)$ for example, we have that the fundamental representation of $\frak{su}(N)$ is the set of traceless complex $N \times N$ matrices obeying $\al + \al^{\dagger} = 0$. The basis of $\frak{su}(N)$ is chosen such that
\beq \label{eq:suNalg}
\[ T^a, T^b \] = f^{abc} T^c \, , \hspace{1cm}  \Tr \( T_{\rm F}^a T_{\rm F}^b \) = -\frac{1}{2}\, \de^{ab} 
\eeq
where the structure coefficients $f^{abc}$ are totally antisymmetric and the ``F" subscript denotes the fundamental representation. The indices of $T^a_{\rm F}$ will be given by greek letters of the end of the alphabet, i.e. the matrix elements are $\(T^a_{\rm F} \)^{\si\ta}$.

For the LLTs we focus on the component connected to the identity ${\rm SO}_1(1,d)$. Since we are also going to consider spinors, the actual group that acts on the $I$ indices is the double cover Spin$(1,d)$. The standard choice of basis of $\frak{spin}(1,d)$ is the one obeying the Lorentz algebra
\beq \label{eq:LorAlg}
\[T^{IJ}, T^{KL}\] = \et^{IL} T^{JK} - \et^{JL} T^{IK} - \et^{IK} T^{JL} + \et^{JK} T^{IL} \, ,
\eeq
where $T^{IJ} \equiv - T^{JI}$. The vector and Dirac representations read
\beq
\(T_{\rm v}^{IJ}\)^K_{\,\,\,L} = \et^{IK}\de^J_L - \et^{IK}\de^J_L \, , \hspace{1cm} T_{\rm D}^{IJ} = \frac{1}{2} \ga^{IJ} \, ,
\eeq
respectively, where $\ga^{I_1 \dots I_p} \equiv \ga^{[I_1} \dots \ga^{I_p]}$ and the $\ga^I$ obey the Clifford algebra $\{ \ga^I, \ga^J \} = 2\, \et^{IJ}$. The following identity will be useful later-on
\beq \label{eq:acomrel}
\ga^I \ga^{JK} = \et^{IK} \ga^J - \et^{IJ} \ga^K + \ga^{IJK} \, .
\eeq
Choosing a representation where $\(\ga^I\)^{\dagger} = \et^{II} \ga^I$ (no summation) and defining the bar conjugation so that it is an involution $\bar{\psi} \equiv \psi^{\dagger} i\ga^0$, we get $\bar{\ga}^I \equiv i \ga^0 \(\ga^I\)^{\dagger} i \ga^0 = - \ga^I$ and thus
\beq \label{eq:barga}
\bar{\ga}^{I_1 \dots I_p} = (-1)^p \ga^{I_p \dots I_1} = (-1)^{p(p+1)/2} \ga^{I_1 \dots I_p} \, .
\eeq
In particular, $\bar{\ga}^{IJ} = - \ga^{IJ}$ so that the Dirac representation of $\frak{spin}(1,d)$ is the set of matrices obeying $\te + \bar{\te} = 0$. We will use greek indices from the beginning of the alphabet for the elements of these matrices, i.e. $\(\ga^{IJ}\)^{\al\be}$.

Let Diff$\( \cal M \)$ denote the group of diffeomorphisms of $\cal M$, i.e. the group of homeomorphisms from $\cal M$ to $\cal M$ that are diffeomorphisms in any local coordinate system, and let ${\rm Diff}_1\( \cal M \)$ denote its component connected to the identity. Every vector field $\xi \in \frak{X}(\cal M)$ generates a one-parameter family of diffeomorphisms $\{ \Xi_t \}_{t \in I \subset \Rs} \subset {\rm Diff}_1( \cal M )$ with $I \ni 0$ defined by
\beq \label{eq:xigen}
\Xi_0 = {\rm id}   \, , \hspace{1cm}   \left. \dot{\Xi}_t \right|_{t = 0} = \xi \, .
\eeq
Inversely, by definition of ${\rm Diff}_1( \cal M )$, every element $\Xi \in {\rm Diff}_1( \cal M )$ sufficiently close to the identity can be though of as the $t = 1$ element of the family generated by some $\xi$. In analogy with the theory of Lie groups of finite dimension, we can thus say that $\frak{X}(\cal M)$ is the ``Lie algebra" of ${\rm Diff}_1( \cal M )$. This is the fundamental representation of ${\rm Diff}_1( \cal M )$, also known as ``passive diffeomorphisms", but the ones of interest here are the tensor representations, also known as ``active diffeomorphisms". In that case the $\Xi \in {\rm Diff}_1( \cal M )$ acts on a section $T$ of a bundle based on $\cal M$ through the pushforward map $T' = \Xi_* T$. Since $\Xi$ is a bijection, we have $\Xi_{*,t} = \Xi^*_{-t}$, where $\Xi^*$ is the pull-back and thus the infinitesimal variation involves the Lie derivative
\beq \label{eq:diffvar}
\de T = - \Lie_{\xi} T \hspace{0.5cm}  \Rightarrow \hspace{0.5cm} \Xi_* = e^{-\Lie_{\xi}}\, .
\eeq
Thus, the generators in that representation are the Lie derivatives and, given the properties of $\Lie$, obey the following algebra
\beq \label{eq:DiffAlg}
\[ \Lie_{\xi}, \Lie_{\xi'} \] = \Lie_{\[ \xi, \xi' \]} \, ,
\eeq
where $\[ \xi, \xi' \]$ is the Lie bracket, the algebraic product of $\frak{X}(\cal M)$. In our global and coordinate-independent setting, there is only one privileged basis for the Lie algebra $\frak{X}(\cal M)$, the one given by the frame vectors $\ep_I$
\beq \label{eq:CLietrans}
\[ \ep_I, \ep_J \] = C_{IJ}^{\,\,\,\,\,\,K} \ep_K \, .
\eeq 
The $C_{IJK} = i_{IJ} \ed e_K$ are known as the ``structure coefficients".

\subsection{Fields and Lagrangians}

We use totally dimensionless units $\hbar = c = 8\pi G = 1$. The fields of the theory we are going to consider are the vielbein, spin connection and YM fields $e^I, \om_{IJ}, A^a \in \Om^1({\cal M})$, respectively, and the Dirac spinor $\psi^{\al\si} \in \Om^0({\cal M}) \otimes \mathbb{G}$, where $\mathbb{G}$ is the set of complex Grassmann numbers. The non-trivial infinitesimal variations under an SU$(N)$ transformation with parameter $\al = \al^a T^a \in \Om^0({\cal M}) \otimes \frak{su}(N)$ are
\beq \label{eq:SUNvar}
\de A^a = -\ced \al^a \equiv -\ed \al^a + f^{abc} \al^b A^c \,   \, , \hspace{0.5cm}  \de \psi = \al^a T^a_{\rm F} \,\psi  \, .
\eeq
Under an LLT with parameter $\te = \frac{1}{2}\, \te_{IJ} T^{IJ} \in \Om^0({\cal M}) \otimes \frak{spin}(1,d)$ we have
\bea 
\de e^I & = & \te^I_{\,\,\,J} e^J \, , \hspace{1cm} \de \psi = \frac{1}{4}\, \te_{IJ} \, \ga^{IJ} \,\psi \nn \\ 
\de \om_{IJ} & = & -\ced \te_{IJ} \equiv - \ed \te_{IJ} + \te_I^{\,\,\,K} \om_{KJ} - \te_J^{\,\,\,K} \om_{KI} \, ,  \label{eq:Spinvar}
\eea
and under a diffeomorphism with parameter $\xi \in \frak{X}$ all fields transform as (\ref{eq:diffvar}). The Lagrangian $D$-form is given by $L \equiv L_{\rm g} + L_{\rm m}$, where $L_{\rm g}$ and $L_{\rm m}$ are the gravitational and matter parts, respectively. For gravity we are going to consider two Lagrangians. The first-order one is the Palatini Lagrangian
\beq \label{eq:LP}
L_{\rm P} \equiv \frac{1}{2} \,\Om_{IJ} \we \ti{e}^{IJ} \, ,
\eeq 
where $\Om_{IJ} \equiv \ed \om_{IJ} + \om_I^{\,\,\,K} \we \om_{KJ}$ are the curvature two-forms of the spin connection. The second-order one is the M\o ller Lagrangian \cite{Mol} (see appendix \ref{sec:GR2})
\beq \label{eq:LM}
L_{\rm M} \equiv -\frac{1}{2} \( F^I_{\,\,\,J} \we \star\, F^J_{\,\,\,I} - \frac{1}{2}\, F \we \star\, F  \) \, ,
\eeq
where $F^{IJ} \equiv e^I \we \ed e^J$ and $F \equiv F^I_{\,\,\,I}$. In the matter sector we have the YM Lagrangian
\beq  \label{eq:LYM2}
L_{\rm YM} \equiv - \frac{1}{2 g^2} \,F^a \we \star\, F^a  \, ,
\eeq
where $F^a = \ed A^a + f^{abc} A^b \we A^c / 2$ are the curvature two-forms of $A^a$, and the Dirac Lagrangian
\bea 
L_{\rm D} & \equiv & {\rm Re} \[ \bar{\psi} \ga^I \ced \psi \] \we \ti{e}_I = \frac{1}{2} \[ \bar{\psi} \ga^I \ced \psi - \ced \bar{\psi} \ga^I \psi \] \we \, \ti{e}_I  \nn \\
 & = &   \frac{1}{2} \[ \bar{\psi} \ga^I \ed \psi - \ed \bar{\psi} \ga^I \psi \] \we \ti{e}_I \nn \\
 & + & \frac{1}{4}\, \om_{IJ} \we \ti{e}_K \, \bar{\psi} \ga^{IJK} \psi +  A^a \we \ti{e}_I \, \bar{\psi} T^a \ga^I  \psi   \, . \label{eq:LD}
\eea
We have used the exterior covariant derivative in the appropriate representations
\beq
\ced \psi \equiv  \( \ed + \frac{1}{4}\, \om_{IJ}\, \ga^{IJ} + A^a T^a  \) \psi \, , \hspace{1cm} \ced \bar{\psi} \equiv  \overline{\ced \psi}  \, ,
\eeq
the definition of complex conjugation on complex Grassmann numbers $\( \psi^{\al\si} \psi^{\be\ta} \)^* = \psi^{\be\ta *}\psi^{\al\si *}$ and (\ref{eq:acomrel}).

\section{Lagrange - Noether formalism} \label{sec:basics}

\subsection{The variational principle}  \label{sec:varprin}

We start by considering a generic field content, i.e. let $\ph^a \in \Om^p (\cal M) \otimes \mathbb{K}$, with $a = 1, \dots, N$ and $\mathbb{K} = \Rs$, $\Cs$ or $\mathbb{G}$, be a set of $N$ $p$-form fields. They are given with a Lagrangian $L = L\( \ph, \ed \ph \) \in \Om^D(\cal M)$ which is local, i.e. the values of $L$ at $p \in {\cal M}$ only depend on the values of $\ph^a$ and $\ed \ph^a$ at that same point. We also ask that $L$ be polynomial in $\ph^a, \ed \ph^a$ except for the vielbein. To our Lagrangian there corresponds an action functional $S \equiv \int_{\cal M} L$ and the variational principle goes as follows. We consider a field configuration $\ph^a$ and an infinitesimal variation $\ph^a \to \ph^a + \de \ph^a$ over $\cal M$. The variation of $\ed \phi^a$ being determined by the one of $\phi^a$, i.e. $\de \ed \ph^a = \ed \de  \ph^a$, the variation of the Lagrangian is given by
\beq \label{eq:Lagvar}
\de L = \de \ph^a \we \frac{\pa L}{\pa \ph^a} + \ed \de  \ph^a \we \frac{\pa L}{\pa \ed \ph^a}   \, .
\eeq
Here the operators $\frac{\pa }{\pa \ph^a}$ and $\frac{\pa }{\pa \ed\ph^a}$ are defined as anti-derivations of degree $-p$ and $-(p+1)$, respectively, satisfying 
\beq
\frac{\pa}{\pa \ph^a}\, \ph^b = \de_a^b \, , \hspace{1cm} \frac{\pa}{\pa \ed \ph^a}\, \ed \ph^b = \de_a^b \, , 
\eeq
they have the same Grassmann degree as $\ph^a$ and we conventionally apply them from the left. The above equations and the fact that they are anti-derivations determine them on all of $\Om ({\cal M} )$. Now the variation of the action being
\beq \label{eq:actvar}
\de S = \int_{\cal M} \de L = \int_{\cal M} \( \de \ph^a \we \frac{\pa L}{\pa \ph^a} + \ed \de  \ph^a \we \frac{\pa L}{\pa \ed \ph^a}  \)   \, ,
\eeq
we integrate by parts the second term and find 
\beq \label{eq:varbtermdrops}
\de S = \int_{\cal M} \de \ph^a \we  \( \frac{\pa L}{\pa \ph^a} - (-1)^p \ed \frac{\pa L}{\pa \ed \ph^a}  \)  + \int_{\pa {\cal M}} \de \ph^a \we \frac{\pa L}{\pa \ed \ph^a}   \, .
\eeq
To get rid of the boundary term we have to restrict to variations such that $\de \ph^a|_{\pa {\cal M}} = 0$. The classical solutions of the theory given by $L$ are the field configurations which make $\de S$ vanish for all such $\de \ph^a$ and therefore obey
\beq \label{eq:EOM}
{\rm EL}_a \equiv \frac{\pa L}{\pa \ph^a} - (-1)^p \ed \frac{\pa L}{\pa \ed \ph^a} = 0 \, .
\eeq
These are the Euler-Lagrange equations in exterior calculus. Note that they imply that the $(D-p)$-form
\beq \label{eq:conscurEOM}
J_a \equiv\frac{\pa L}{\pa \ph^a}
\eeq
is exact when evaluated on a classical solution and thus $\ed J_a = 0$. If $p = 0$, then this is a trivial identity since $J_a \in \Om^D(\cal M)$, but if $p > 0$, then this is a genuine conservation equation, valid on classical solutions. It is therefore not surprising that it will have a direct relation with the Noether currents in the case where $p =1$, as we will see later on. For later reference, we will call these $d$-forms $J_a$ the ``Euler-Lagrange", or simply, ``EL" currents.

\subsubsection{Example}   \label{sec:EOMex}

Let us consider the Palatini Lagrangian for gravity here. The equation of motion of the vielbein is
\beq  \label{eq:EOMe}
{\rm EL}_I \equiv \frac{\pa L}{\pa e^I} + \ed \frac{\pa L}{\pa \ed e^I} =  -G_I + T_I = 0 \, ,
\eeq
where
\beq \label{eq:Gdef}
G_I \equiv  -\frac{\pa L_{\rm P}}{\pa e^I} - \ed \frac{\pa L_{\rm P}}{\pa \ed e^I} \os{(\ref{eq:binprod})}{=} -\frac{1}{2}\,\ti{e}_{IJK} \we \Om^{JK} \, ,
\eeq
is the Einstein tensor in first-order vielbein GR and
\bea \label{eq:GREMT}
T_I & \equiv & \frac{\pa L_{\rm m}}{\pa e^I} + \ed \frac{\pa L_{\rm m}}{\pa \ed e^I} \nn \\
 & \os{(\ref{eq:binprod})(\ref{eq:starvar})(\ref{eq:inprodsym})}{=} & \frac{1}{2 g^2} \( i_I F^a \we \star\, F^a - F^a \we i_I \star F^a  \) \nn \\
 & & - {\rm Re} \[ \bar{\psi} \ga^J \ced \psi \] \we \ti{e}_{JI} \, ,
\eea
is the standard definition of the matter energy-momentum tensor in GR, i.e. the variational derivative of the matter action with respect to the gravitational field. The equation of motion of the spin connection is
\bea
{\rm EL}_{IJ} & \equiv & \frac{\pa L}{\pa \om^{IJ}} + \ed \frac{\pa L}{\pa \ed \om^{IJ}} = \frac{1}{2}\, \ced \ti{e}_{IJ} + \frac{1}{4} \, \bar{\psi} \ga_{IJK} \psi \, \ti{e}^K \nn  \\
 & \os{(\ref{eq:bexder})}{=} & \frac{1}{2}\, \ti{e}_{IJK} \we \Te^K + \frac{1}{4} \, \bar{\psi} \ga_{IJK} \psi \, \ti{e}^K = 0 \, , \label{eq:EOMom}
\eea
where $\Te^I \equiv \ed e^I + \om^I_{\,\,\,J} \we e^J$ are the torsion two-forms. The equation of motion of $A^a$ gives
\bea
{\rm EL}^a & \equiv & \frac{\pa L}{\pa A^a} + \ed \frac{\pa L}{\pa \ed A^a} \nn \\
 & \os{(\ref{eq:inprodsym})}{=} & - \frac{1}{g^2} \[ \frac{\pa F^b}{\pa A^a} \we \star\, F^b  + \ed \( \frac{\pa F^b}{\pa \ed A^a} \we \star\, F^b\) \] \nn \\
 & & + \ti{e}_I \, \bar{\psi} T^a \ga^I  \psi \nn \\
 & = &  -  \frac{1}{g^2} \[ \ed \star F^a + f^{abc} A^b \we \star\, F^c \] + \ti{e}_I \, \bar{\psi} T^a \ga^I  \psi \nn \\
 & \equiv & - \frac{1}{g^2} \, \ced \star F^a + \ti{e}_I \, \bar{\psi} T^a \ga^I  \psi = 0 \, . \label{eq:EOMA}
\eea
Finally, the equation of motion of $\bar{\psi}^{\al\si}$ is
\bea
{\rm EL}^{\al\si} & \equiv & \frac{\pa L}{\pa \bar{\psi}^{\al\si}} - \ed \frac{\pa L}{\pa \ed \bar{\psi}^{\al\si}}  \\
 & \os{(\ref{eq:bexder})(\ref{eq:bdbprod})}{=} & \[\ga^I \ed \psi + \frac{1}{2}\, \hat{\om}^I_{\,\,\,J} \ga^J \psi  \right. \nn \\
 & & \left. + \frac{1}{4}\, \om_{JK}\, \ga^{IJK} \psi + A^a T^a \psi \]^{\al\si} \we \ti{e}_I   \nn \\
 & \os{(\ref{eq:acomrel})}{=} & \[ \ga^I \hat{\ced} \psi + \frac{1}{4} \, K_{JK} \ga^{IJK} \psi \]^{\al\si} \we \ti{e}_I = 0 \, , \nn
\eea
where we have used (\ref{eq:acomrel}), $\hat{\ced}$ is the covariant derivative using the Levi-Civita spin connection $\hat{\om}[e]$, i.e. $\hat{\ced} e^I = 0$, and $K_{IJ} \equiv \om_{IJ} - \hat{\om}_{IJ}$ are the contorsion $2$-forms. We have thus retrieved the Palatini, Yang-Mills and Dirac equations of motion from a variational principle without ever having to introduce coordinates, only some simple identities of appendix \ref{sec:useide}. Finally, we have three non-zero form fields and the corresponding EL currents are
\bea
J_I & \equiv & \frac{\pa L}{\pa e^I} = {\rm EL}_I \, , \\
J^{IJ} & \equiv & \frac{\pa L}{\pa \om_{IJ}} = \om^{[I|}_{\,\,\,\,K} \we \ti{e}^{K|J]} + \frac{1}{4} \, \bar{\psi} \ga^{IJK} \psi \, \ti{e}_K \, ,\\
J^a & \equiv & \frac{\pa L}{\pa A^a} = -\frac{1}{g^2}\, f^{abc} A^b \we \star\, F^c + \ti{e}_I \,\bar{\psi} T^a \ga^I  \psi  \, .
\eea
Note that the one of $e^I$ is zero on-shell since the Lagrangian we chose does not depend on $\ed e^I$. 

Finally, note that equation (\ref{eq:EOMom}) is equivalent to $\Te^I = -\frac{1}{4} \, \bar{\psi} \ga^{IJK} \psi \, e_{JK}$ which implies $\om_{IJ} = \hat{\om}_{IJ}[e] + \frac{1}{4}\, \bar{\psi} \ga_{IJK} \psi \, e^K$. We thus retrieve the well-known result that $L_{\rm P}$ is classically equivalent to the Einstein-Hilbert theory only if there is no matter coupling to $\om^{IJ}$.

  %and one finds that $L_{\rm P}$ is classically equivalent to the Einstein-Hilbert theory only if there is no matter coupling to $\om^{IJ}$, \footnote{Indeed, multiplying by $e^L$, using (\ref{eq:binprod}), taking the Hodge dual and using (\ref{eq:idualtie}) we obtain $\Te^{IJL} + 2\, \et^{L[I} \Te^{J]K}_{\,\,\,\,\,\,\,\,K} = \frac{1}{2} \, \bar{\psi} \ga^{IJL} \psi$, where $\Te_{IJK} \equiv i_{IJ} \Te_K$. Any trace of the latter is zero on-shell so $\Te^I = -\frac{1}{4} \, \bar{\psi} \ga^{IJK} \psi \, e_{JK}$ which implies $\om_{IJ} = \hat{\om}_{IJ}[e] + \frac{1}{4}\, \bar{\psi} \ga_{IJK} \psi \, e^K$.}.

\subsection{Noether's theorem}

\subsubsection{Currents} \label{sec:NoetherCurr}

A continuous symmetry is a transformation
%\footnote{Here by ``active transformation" we mean a map sending a section, in fibre bundle language, to another section.}
 of the fields under a continuous group whose infinitesimal version makes the Lagrangian transform as
\beq \label{eq:Lsymvar}
\de L = - \ed K \, ,
\eeq
for some $d$-form $K$. Equivalently, the transformation is a symmetry if the action is invariant up to a boundary term. Noether's theorem states that to every such symmetry there corresponds a $d$-form ${\cal J}$, the Noether current, which is conserved when evaluated on classical solutions, i.e. we have an identity $\ed {\cal J} \sim {\rm EL}$. In field theory one usually works with the dual current $1$-form $j \sim \star\, {\cal J} \in \Om^1(\cal M)$, for which the above equation is expressed in terms of the codifferential $\bar{\ed} j \equiv \na^{\mu} j_{\mu} \sim {\rm EL}$. It is however the $d$-form ${\cal J}$ which is the natural coordinate-independent representation of a ``current", since the latter must be integrated over a $d$-volume in order to give a charge. Even the conservation equation is simpler in terms of $\cal J$ since $\ed$ is an anti-derivation while $\bar{\ed}$ is not. 

To determine $\cal J$, and thus prove the theorem, we compute the infinitesimal variation of the Lagrangian but as induced by the variation of the fields
\bea
\de L & = & \de \ph^a \we \frac{\pa L}{\pa \ph^a} + \ed \de \ph^a \we \frac{\pa L}{\pa \ed \ph^a} \nn \\
 & \os{\(\ref{eq:EOM}\)}{=} & \de \ph^a \we {\rm EL}_a + (-1)^p \de \ph^a \we \ed \frac{\pa L}{\pa \ed \ph^a}  + \ed \de \ph^a \we \frac{\pa L}{\pa \ed \ph^a}   \nn \\
 & = & \de \ph^a \we {\rm EL}_a  + \ed \( \de \ph^a \we \frac{\pa L}{\pa \ed \ph^a} \)  \, .
\eea
Equating this with (\ref{eq:Lsymvar}) and defining the $d$-form
\beq \label{eq:JNoether}
{\cal J} \equiv \de \ph^a \we \frac{\pa L}{\pa \ed \ph^a} + K \, ,
\eeq
one gets the desired identity $\ed {\cal J} = -\de \ph^a \we {\rm EL}_a$. Eq. (\ref{eq:JNoether}) is therefore the definition of the Noether current. By the Poincar\'e lemma, if $\ed {\cal J} = 0$ on-shell, then there exists locally a $(d-1)$-form ${\cal U}$, called the ``superpotential", such that ${\cal J} = \ed {\cal U}$. In the following examples we will see that, as a consequence of assuming the classical solutions to be global, this ${\cal U}$ form also exists globally.

\subsubsection{Example}

We consider the Noether currents associated with the SU$(N)$ and Spin$(1,d)$ transformations. We will use a shorthand notation $\pi^a \equiv -g^{-2} \star F^a$ to simplify our expressions. The variations are given in (\ref{eq:SUNvar}) and (\ref{eq:Spinvar}) and the Lagrangian is invariant so $K = 0$. Thus, the currents associated to the transformations generated by $\al = \al^a T^a \in \Om^0({\cal M}) \otimes \frak{su}(N)$ and $\te = \frac{1}{4}\, \te_{IJ} \ga^{IJ} \in \Om^0({\cal M}) \otimes \frak{spin}(1,d)$ are
\bea  
{\cal J}[\al] & \equiv & \de A^a \we \frac{\pa L}{\pa \ed A^a} + \de \psi^{\al\si} \, \frac{\pa L}{\pa \ed \psi^{\al\si}} + \de \bar{\psi}^{\al\si} \, \frac{\pa L}{\pa \ed \bar{\psi}^{\al\si}} \nn \\
 & = & \( \ed \al^a + f^{abc} A^b \, \al^c \) \we \pi^a + \al^a\,\ti{e}_I \,\bar{\psi} T^a \ga^I  \psi \, , \label{eq:SUNcurr} 
\eea
and
\bea
{\cal S}[\te] & = & \de e^I \we \frac{\pa L}{\pa \ed e^I} + \de \om_{IJ} \we \frac{\pa L}{\pa \ed \om_{IJ}} \\
 & & + \de \psi^{\al\si} \, \frac{\pa L}{\pa \ed \psi^{\al\si}} + \de \bar{\psi}^{\al\si} \, \frac{\pa L}{\pa \ed \bar{\psi}^{\al\si}} \nn \\
 & = & -\frac{1}{2} \( \ed \te_{IJ} - 2\,\te_I^{\,\,\,K} \om_{KJ} \) \we \ti{e}^{IJ} + \frac{1}{4}\, \te_{IJ} \ti{e}_K \, \bar{\psi} \ga^{IJK} \psi \, ,  \nn \label{eq:Spincurr}
\eea
respectively. The fact that these currents are conserved on-shell for any $\al$ and $\te$ is the mark of the redundancy in the apparent number of degrees of freedom in a gauge theory. We then consider the special cases
\bea
{\cal J}^a & \equiv & {\cal J}\[T^a\] =  f^{abc} A^b \we \pi^c + \ti{e}_I \,\bar{\psi} T^a \ga^I  \psi = J^a \, ,  \\
{\cal S}^{IJ} & \equiv & {\cal S}\[ T^{IJ} \] = \om^{[I|}_{\,\,\,\,K} \we \ti{e}^{K|J]} + \frac{1}{4}\, \ti{e}_K \, \bar{\psi} \ga^{IJK} \psi = J^{IJ}\, . \nn \label{eq:SpincurrIJ}
\eea
As anticipated earlier, the EL currents $J^a \equiv \frac{\pa L}{\pa A^a}$ and $J^{IJ} \equiv \frac{\pa L}{\pa \om_{IJ}}$ are thus nothing but a special subset of the Noether currents associated with the groups $A$ and $\om$ gauge, respectively. Moreover, these currents are the ones one obtains in the case where the gauge field is absent and the symmetry is only global. Here we see that by gauging the symmetry we obtain a whole lot of new conserved currents that are indexed by geometric objects, the fields $\al$ and $\te$ in the adjoint representation of their respective groups. 
 
Now, all of these currents are not independent but can be expressed in terms of $J^a$ and $J^{IJ}$ using (\ref{eq:SUNcurr}) and (\ref{eq:Spincurr})
\beq  \label{eq:intcurrdecomp}
{\cal J}[\al] = \al^a J^a - \ed \al^a \we \pi^a \, , \hspace{0.5cm} {\cal S}[\te] = \te_{IJ} J^{IJ} - \frac{1}{2} \, \ed \te_{IJ} \we \ti{e}^{IJ}
\eeq
Using the equations of motion $J^a = -\ed \pi^a$ and $J^{IJ} = -\frac{1}{2}\,\ed \ti{e}^{IJ}$, we get that on classical solutions
\bea 
\left. {\cal J}[\al] \right|_{{\rm EL} = 0} & = & -\al^a \ed \pi^a - \ed \al^a \we \pi^a = -\ed \( \al^a \pi^a \)  \, , \\  \label{eq:intcurrexact}
\left. {\cal S}[\te] \right|_{{\rm EL} = 0} & = & -\frac{1}{2} \( \te_{IJ} \ed \ti{e}^{IJ} + \ed \te_{IJ} \we \ti{e}^{IJ}  \) = -\frac{1}{2}\,\ed \( \te_{IJ} \ti{e}^{IJ} \)  \, . \nn
\eea
So the superpotentials exist globally indeed. It is important to understand the difference between ${\cal J}[\al]$ and ${\cal J}^a$. The former is the Noether current associated to a covariant field $\al$ and is thus gauge-invariant, as can be seen in (\ref{eq:SUNcurr}). On the other hand, the EL current ${\cal J}^a$ corresponds to the fixed choice $\al = T^a$ and thus transforms inhomogeneously. Note finally that ${\cal J}[\al]$ is $\Rs$-linear in its argument
\beq
{\cal J}[\al + \be] =  {\cal J}[\al] + {\cal J}[\be] \, , \hspace{1cm}  {\cal J}[c\, \al] = c \,{\cal J}[\al] \, ,  
\eeq
for all $c \in \Rs$. Of course, what has been said for ${\cal J}[\al]$ in this paragraph holds analogously for ${\cal S}[\te]$ as well.

\subsubsection{Charges} \label{sec:NoetherCharges}

Consider now a Noether current ${\cal J}$ evaluated on a given field configuration $\ph^a$ which is not necessarily a classical solution. We can construct the corresponding Noether charge $Q$ contained in a region $\Si \in \De^d$, that is, define a map $Q : \De^d \to \Rs$
\beq
Q\( \Si \) \equiv \int_\Si {\cal J} \, .
\eeq
Let us now express the conservation law in terms of $Q$. We start by choosing a local evolution direction, that is, a vector field $\xi \in \frak{X}({\cal M})$. We then consider the one-parameter subgroup $\{ \Xi_t \}_{t \in I \subset \Rs}$ of ${\rm Diff}_1(\cal M)$ that it generates, i.e. the one satisfying (\ref{eq:xigen}). We also consider the continuous one-parameter family of submanifolds $\Si_t \equiv \Xi_t(\Si)$, the corresponding charges $Q(t) \equiv Q( \Si_t )$ and also define the ``tube"
\beq
W \equiv \bigcup_{t \in I} \Si_t \, .
\eeq
Thus, $Q(t)$ is the charge contained in the $\Si_t$ hypersurface and the latter evolves along the flow-lines of $\xi$. The variation of $Q(t)$ with respect to $t$ gives   
\bea
\dot{Q}(t) & \equiv & \lim_{\ep \to 0} \frac{1}{\ep} \[ Q ( \Si_{t+\ep} ) - Q( \Si_t ) \] \nn \\
 & = & \lim_{\ep \to 0} \frac{1}{\ep} \[ \int_{\Xi_{t+\ep}(\Si )} {\cal J} - \int_{\Xi_t (\Si )} {\cal J} \] \nn \\
 & = & \int_\Si \lim_{\ep \to 0} \frac{1}{\ep}  \[ \Xi_{t+\ep}^* {\cal J}  - \Xi_t^* {\cal J} \] \equiv \int_\Si \Xi_t^* \Lie_{\xi} {\cal J} = \int_{\Si_t} \Lie_{\xi} {\cal J}  \nn \\
 & = & \int_{\Si_t} \( i_{\xi} \ed + \ed i_{\xi} \) {\cal J} = \int_{\Si_t} i_{\xi} \ed {\cal J} + \int_{\pa {\Si_t}} i_{\xi} {\cal J}    \, , 
\eea 
where $\Xi_t^*$ is the pullback with respect to $\Xi_t$ and we have used the definition of the Lie derivative as the generator of pullbacks. Focusing from now on on solutions of the equations of motion for $\ph^a$, the first term drops by current conservation and we are left with
\beq \label{eq:NoetherQ}
\left. \dot{Q}(t) \right|_{{\rm EL} = 0} = \int_{\pa {\Si_t}} i_{\xi} {\cal J} \, ,
\eeq 
which is the Noether conservation law in terms of the charge: the variation of the charge within $\Si_t$ is entirely determined by the current normal to $\xi$ at the boundary. More precisely, the vector field $\xi$ determines the shape of the boundary of $W$ and $i_{\xi} {\cal J} \sim \star \( \xi^{\flat} \we j \)$. Therefore, this expression captures the part of $j$ which is normal to $\pa W$. The standard use of this law is in the case where the $\Si_t$ are the space-like leaves of a foliation and $\xi$ is the time-like coordinate-induced vector field $\pa_t$. It then reduces to the fact that the variation of the charge in time inside the space-region $\Si_t$ is equal to the integrated current flux through the boundary of $W$ at the level $t$. This takes both into account the flow of the current out of the volume and the fact that the volume itself may vary with time. Here we see the full reach of Noether's theorem since it applies to {\it any} vector-field $\xi$ and hypersurface $\Si$, whether the corresponding family $\Si_t$ is a foliation or not, and independently of its space-time interpretation. 

As shown in the example given above, in the case of gauge symmetries the superpotential ${\cal U}$ is a local function of the fields {\it and} their derivatives, and the charge of on-shell configurations can be written
\beq
\left. Q(\Si)  \right|_{{\rm EL} = 0} = \int_{\pa \Si} {\cal U} \, .
\eeq
Note that it is crucial that $\cal U$ depends on the derivative of the gauge field, since this makes $Q(\Si)$ sensitive to the fields in an infinitesimally thick but still $d$-dimensional region around $\pa \Si$. Therefore, the total charge can only be computed by starting with a compact $\Si$ and then sending the boundary to infinity. If $Q$ depended only on the field values on $\pa \Si$, then vanishing boundary conditions at infinity would simply imply zero total charges.

\section{The Noether currents of diffeomorphisms}  \label{sec:diffNo}

Let us now consider the Noether current associated to diffeomorphisms. All fields transform infinitesimally as
\beq 
\de \ph^a = - \Lie_{\xi} \ph^a = - i_{\xi} \ed \ph^a - \ed i_{\xi} \ph^a \, .
\eeq
General covariance of the theory implies that the Lagrangian is a tensor, and more specifically a $D$-form, so $\ed L = 0$ and thus $\de L = -\ed i_{\xi} L$, which is a total derivative and therefore we have a symmetry. Following the derivation of Noether's theorem, we have $K = i_{\xi} L$ and $\de \ph^a = -\Lie_{\xi} \ph^a$, so the Noether current associated to the diffeomorphism in the $\xi$ direction is 
\beq \label{eq:diffcurr1}
{\cal P}[\xi] = -\Lie_{\xi} \ph^a  \we \frac{\pa L}{\pa \ed \ph^a} + i_{\xi}L \, .
\eeq
This is a generalization of (minus) the covariant Hamiltonian of ref\cite{NesCovHam} which corresponds to the case $\xi = \pa_t$. Just like in the case of internal symmetries, we can compute the current corresponding to the basis elements of the Lie algebra $\frak{X}(\cal M)$. If we expect to relate this object to the EL current of the vielbein, by analogy with the internal symmetries, we must give it an $I$ index, so we evaluate it on the $\ep_I$ basis
\beq \label{eq:NEM}
{\cal P}_I \equiv {\cal P}[\ep_I] = -\Lie_I \ph^a  \we \frac{\pa L}{\pa \ed \ph^a} + i_I L \, .
\eeq
The problem with this expression is that, unlike in the case of internal symmetries, the EL current of $e^I$ is not a Noether current ${\cal P}_I \neq \frac{\pa L}{\pa e^I}$. This can be seen in full generality by noting that the only non-trivial step in computing the latter is when $\frac{\pa}{\pa e^I}$ acts on $\star$, whose solution is given by (\ref{eq:starvar}). This cannot produce a term $\sim \ed i_I \ph^a$ which is present in (\ref{eq:NEM}) through $\Lie_I \ph^a$ for forms of non-zero degree. Moreover, if we express ${\cal P}[\xi]$ in terms of the ${\cal P}_I$ using (\ref{eq:diffcurr1})
\beq \label{eq:Diffcurrdecomp}
{\cal P}[\xi] = \xi^I {\cal P}_I - \ed\xi^I \we i_I \ph^a \we \frac{\pa L}{\pa \ed \ph^a} \, ,
\eeq
and compare with (\ref{eq:intcurrdecomp}), we see that all non-zero-form fields appear in the second term. Thus, we cannot yet deduce the superpotential corresponding to ${\cal P}[\xi]$. This issue is addressed in the following section.

\subsection{The anholonomic representation}

The problem raised above is related to the fact that, because of the existence of a vielbein, an ambiguity arises regarding precisely non-zero forms. Should one take the $p$-forms $\ph^a$ as the independent fields, or should one rather consider their components in the vielbein basis $\ph^a_{I_1\dots I_p} \equiv i_{I_p \dots I_1} \ph^a \in \Om^0$? We will call the first case the ``holonomic representation" while the second one will be the ``anholonomic representation", since the corresponding vector bases $\pa_{\mu}$ and $\ep_I$ have trivial and non-trivial Lie brackets, respectively. Using $\ph^a$ to denote all fields {\it but} the vielbein from now on, we have the two theories 
\beq
L^{\rm hol} \equiv L^{\rm hol} \[\phi^a,  e^I \]  \, , \hspace{1cm}  L^{\rm an} \equiv L^{\rm an} \[ \ph^a_{I_1 \dots I_p},  e^I \] \, .
\eeq
The holonomic representation is well suited to gauge theory since the EL currents and holonomies make use of the gauge field $1$-forms, not the Lorentz-indexed $0$-forms. Moreover, the gauge transformations of the gauge fields in terms of the $0$-forms make use of the inverse vielbein, so they are less natural. As we will show now, the anholonomic representation, however, is well suited for gravity because then the vielbein formally behaves as the gauge field associated to diffeomorphisms, i.e. in total analogy with the properties of the gauge fields we have seen until now. 

The first thing to show of course, is that both representations are classically equivalent, i.e. that their equations of motion imply one another. This is a priori obvious since the two choices of independent fields are related by a non-linear but bijective field redefinition,
\beq
\ph^a = \frac{1}{p!} \, \ph^a_{I_1 \dots I_p} \, e^{I_1 \dots I_p} \, , \hspace{0.5cm}  \ph^a_{I_1 \dots I_p} = i_{I_p \dots I_1} \ph^a \, .
\eeq
We show the equivalence explicitly in appendix \ref{sec:stdsveq} because, in the process, we see that one can choose a mixed representation for the equations of motion. We can compute them in the anholonomic one for the vielbein and in the holonomic one for the rest of the fields. We also show that the Noether currents are the same, even though some fields have changed Spin$(1,d)$ and ${\rm Diff}_1({\cal M})$ representations. So let us consider the $L^{\rm an}$ theory, and for notational simplicity let us absorb the $I$ indices in the generic internal index $a$, i.e. let us write $\ph^{\ti{a}} \equiv \ph_{I_1 \dots I_p}^a$ and keep in mind that now $\ph^{\ti{a}} \in \Om^0({\cal M})$. The key property of the anholonomic representation is that the Lie derivative with respect to $\ep_I$ becomes simply
\beq \label{eq:Liesimp}
\left. \Lie_I \right|_{\rm an} = i_I \ed  \, ,
\eeq
on all fundamental fields, since $i_I \ph^{\ti{a}} = 0$ and $i_I e^J = \de_I^J$. Considering the $\ep_I$ basis for $\frak{X}(\cal M)$ means that we take $\Lie_I = i_I \ed$ as our generators in the active representation of ${\rm Diff}_1(\cal M)$. These are nothing but the $\ep_I$ vectors seen as derivations. Then, under an infinitesimal diffeomorphism the $0$-forms transform homogeneously
\beq
\de_{\xi} \ph^{\ti{a}} = - \Lie_{\xi} \ph^{\ti{a}} = -\xi^I \Lie_I \ph^{\ti{a}} \, ,
\eeq
while the vielbein is the only field transforming ``inhomogeneously"
\beq
\de_{\xi} e^I = - \Lie_{\xi} e^I  = -  \ed \xi^I - \xi^J \Lie_J e^I \, .
\eeq
Note the analogy with the transformation of $A^a$ in (\ref{eq:SUNvar}) and $\om_{IJ}$ in (\ref{eq:Spinvar}). We have the same inhomogenous part, while the homogeneous one is the Lie derivative in different respective senses. Here it is the Lie derivative with respect to the base manifold $\cal M$, while for the internal transformations it is the Lie derivative with respect to the SU$(N)$ and Spin$(1,d)$ fibres\footnote{Indeed, in those cases one can write the transformations in terms of the algebra-valued fields $A \equiv A^a T^a \in \Om^1({\cal M}) \otimes \frak{su}(N)$ and $\om \equiv \frac{1}{2} \, \om_{IJ} T^{IJ}\in \Om^1({\cal M}) \otimes \frak{spin}(1,d)$ where it reads
\beq
\de A = - \ced \al = -\ed \al - \[ A, \al \] \, , \hspace{1cm}  \de \om = - \ced \te = -\ed \te - \[ \om, \te \] \, , 
\eeq
and the commutator with $\al, \te$ is nothing but the Lie derivative in the $\al, \te$ directions, seen as left-invariant vector fields, on their respective group manifolds.}. Thus, the vielbein formally transforms as the gauge field associated to ${\rm Diff}_1({\cal M})$. The Noether currents are the same but are now computed using $L^{\rm an}$, so (\ref{eq:NEM}) now reads
\beq \label{eq:simpEMtens}
{\cal P}_I = - i_I \ed e^J \we \frac{\pa L^{\rm an}}{\pa \ed e^J} - i_I \ed \ph^{\ti{a}} \we \frac{\pa L^{\rm an}}{\pa \ed \ph^{\ti{a}}} + i_I L^{\rm an} \, ,
\eeq
and the general Noether current associated to the diffeomorphism in the $\xi$ direction (\ref{eq:Diffcurrdecomp}) is
\beq \label{eq:Pdecomp}
{\cal P}[\xi] = \xi^I {\cal P}_I - \ed\xi^I \we \frac{\pa L^{\rm an}}{\pa \ed e^I} \, .
\eeq
This result is again analogous to the case of the internal symmetries, see (\ref{eq:intcurrdecomp}). 

The antiderivations $i_I$ and $\frac{\pa}{\pa e^I}$ now have a lot in common. Since they are equal on vielbein products, and now all fields have been decomposed into linear combinations of vielbein products, they are also equal when acting on any combination of the fields which does not contain derivatives, like the potential part of $L^{\rm an}$ for instance. For forms containing derivatives however they do differ since
\beq
\frac{\pa}{\pa e^I} \ed \ph^{\ti{a}} = 0 \neq i_I \ed \ph^{\ti{a}} \, .
\eeq
Thus, acting with $i_I$ on $L^{\rm an}$ we get $\frac{\pa L^{\rm an}}{\pa e^I}$ plus terms $\sim i_I\ed e^J$ and $\sim i_I\ed \ph^{\ti{a}}$. Since $L^{\rm an}$ is polynomial in $\ed e^I$ and $\ed \ph^{\ti{a}}$, the extra term which is not captured by $\frac{\pa }{\pa e^I}$ is simply
\beq
i_I L^{\rm an} - \frac{\pa L^{\rm an}}{\pa e^I} = i_I \ed e^J \we \frac{\pa L^{\rm an}}{\pa \ed e^J} + i_I \ed \ph^{\ti{a}} \we \frac{\pa L^{\rm an}}{\pa \ed \ph^{\ti{a}}} \, .
\eeq
Now, isolating $\frac{\pa L^{\rm an}}{\pa e^I}$ and using (\ref{eq:simpEMtens}), we get
\beq \label{eq:NeqT}
{\cal P}_I = \frac{\pa L^{\rm an}}{\pa e^I} \, ,
\eeq
which is again analogous to the case of internal symmetries: the EL current of the vielbein $J^{\rm an}_I \equiv \frac{\pa L^{\rm an}}{\pa e^I}$ is nothing but a special case of the Noether currents associated to the group $e^I$ gauges. The only peculiarity is that one has to go to the anholonomic representation for this to hold. We finish the comparison with gauge theory by taking (\ref{eq:Diffcurrdecomp}), using (\ref{eq:NeqT}) and evaluating everything on classical solutions $J^{\rm an}_I = -\ed \frac{\pa L^{\rm an}}{\pa \ed e^I}$ to get the analogue of (\ref{eq:intcurrexact})
\bea 
\left. {\cal P}[\xi] \right|_{{\rm EL} = 0} & = & -\xi^I \ed \frac{\pa L^{\rm an}}{\pa \ed e^I} - \ed\xi^I \we \frac{\pa L^{\rm an}}{\pa \ed e^I} \nn \\
 & = & -\ed \( \xi^I \frac{\pa L^{\rm an}}{\pa \ed e^I} \)  \, . \label{eq:Diffcurrexact}
\eea
Thus, the ${\cal P}[\xi]$ too are globally exact when on-shell and we can now compute the superpotential straightforwardly.

\subsubsection{Example}

We illustrate the equality ${\cal P}_I = \frac{\pa L^{\rm an}}{\pa e^I}$ with the YM Lagrangian. Computed for example in the holonomic representation
\bea \label{eq:EMTYM2std}
{\cal P}^{\rm YM}_I & \os{(\ref{eq:JNoether})}{\equiv} & -\Lie_I A^a \we \frac{\pa L_{\rm YM}^{\rm hol}}{\pa \ed A^a} + i_I L_{\rm YM}^{\rm hol} \nn \\
 & \os{(\ref{eq:inprodsym})}{=} & \frac{1}{g^2} \[ \Lie_I A^a \we \frac{\pa F^b}{\pa \ed A^a} \we \star \, F^b - \frac{1}{2} \, i_I \( F^a \we \star\, F^a \) \] \nn \\
 & = & \frac{1}{g^2} \[ \frac{1}{2}\, i_I F^a \we \star\, F^a - \frac{1}{2}\, F^a \we i_I \star F^a \right. \nn \\
 & & \left. + \( \ed A^a_I + f^{abc} A^b A^c_I \) \we \star\, F^a \] \, ,
\eea
where $A_I^a \equiv i_I A^a$. To show the equality, we first need the curvature $F^a$ in terms of $A^a_I$
\bea
F^a[A_I, e^I] = \ed A_I^a \we e^I + A_I^a \ed e^I + \frac{1}{2}\, f^{abc} A_I^b A_J^c \, e^{IJ} 
\eea  
and then we get
\bea
\frac{\pa L_{\rm YM}^{\rm scl}}{\pa e^I} & \os{(\ref{eq:inprodsym})}{=} & -\frac{1}{g^2} \,  \frac{\pa F^a[A_I, e^I]}{\pa e^I} \we \star\, F^a \nn \\
 & & - \frac{1}{2g^2} \,  F^a \we \( \frac{\pa}{\pa e^I} \,\star \) F^a   \\
 & \os{(\ref{eq:starvar})}{=} & \frac{1}{g^2} \[  \( \ed A^a_I + f^{abc} A^b A^c_I \) \we \star\, F^a  \right. \nn \\
 & & \left. + \frac{1}{2}\, i_I F^a \we \star\, F^a - \frac{1}{2}\, F^a \we i_I \star F^a \] = {\cal P}^{\rm YM}_I \, . \nn
\eea

\section{Energy-momentum} \label{sec:EM}

\subsection{Relation to the GR matter energy-momentum tensor}

Let us compute the general relation between the two definitions of the matter energy-momentum tensor that are the Noether current one
\beq
{\cal P}^{\rm m}_I = \frac{\pa L_{\rm m}^{\rm an}}{\pa e^I}
\eeq
and the one of GR, for which it must be noted that it is computed in the holonomic representation
\beq
T_I \equiv \frac{\de S_{\rm m}^{\rm hol}}{\de e^I} \, .
\eeq
We start with the latter and perform a computation analogous to (\ref{eq:ELdif}) for the matter sector only
\bea
T_I & = & \frac{\de S_{\rm m}^{\rm hol}}{\de e^I} = \frac{\de S_{\rm m}^{\rm an}}{\de e^I} - i_I \ph^a \we {\rm EL}^{\rm hol}_a \nn \\
 & = & \frac{\pa L_{\rm m}^{\rm an}}{\pa e^I} + \ed \frac{\pa L_{\rm m}^{\rm an}}{\pa \ed e^I} - i_I \ph^a \we {\rm EL}^{\rm hol}_a \nn \\
 & = & {\cal P}^{\rm m}_I + \ed \frac{\pa L_{\rm m}^{\rm an}}{\pa \ed e^I} - i_I \ph^a \we {\rm EL}^{\rm hol}_a \, ,  \label{eq:TPrel}
\eea
i.e. here the $\ph^a$ are all fields but the vielbein and the spin connection, in the standard representation. Thus, the two energy-momentum tensors are related by a term which is a total derivative when the matter fields are on-shell. Note that the relation is fully general, i.e. it applies to any matter action.

The advantage of the GR definition is that it is fully covariant, as a consequence of being the variation of a fully invariant action. This is not the case of the Noether currents as we show here using the Palatini Lagrangian for gravity in the holonomic representation
\bea 
{\cal P}[\xi] & = & -\Lie_{\xi} e^I  \we \frac{\pa L}{\pa \ed e^I} -\Lie_{\xi} \om^{IJ}  \we \frac{\pa L}{\pa \ed \om^{IJ}} \nn \\
 & & -\Lie_{\xi} A^a  \we \frac{\pa L}{\pa \ed A^a} -\Lie_{\xi} \psi^{\al\si}  \we \frac{\pa L}{\pa \ed \psi^{\al\si}}  \nn \\
 & &  -\Lie_{\xi} \bar{\psi}^{\al\si}  \we \frac{\pa L}{\pa \ed \bar{\psi}^{\al\si}}  + i_{\xi}L\nn \\
 & \os{(\ref{eq:SUNcurr})(\ref{eq:Spincurr})}{=} & \( T_I - G_I \) \xi^I + {\cal J}[i_{\xi} A] + {\cal S}[i_{\xi}\om]  \, . \label{eq:Pex}
\eea 
The first two terms are the matter energy momentum tensor as defined in GR and the Einstein tensor (\ref{eq:Gdef}), in the combination which vanishes on shell. The rest can be expressed as the Noether currents of the internal symmetries evaluated on the $\xi$-projection of the corresponding gauge fields. Therefore, ${\cal P}[\xi]$ is invariant only under internal gauge transformations whose parameters obey $\Lie_{\xi} \al^a = 0$ and $\Lie_{\xi} \te_{IJ} = 0$. Considering more general transformations changes ${\cal P}[\xi]$, the most extreme examples being the transformations that bring the fields to the generalized Weyl gauge $i_{\xi} \om_{IJ} = 0$ and $i_{\xi} A^a = 0$, in which case ${\cal P}[\xi] = 0$ when on-shell since $G_I = T_I$.

\subsection{Gravitational energy-momentum complexes}

We now show how one can obtain some known energy-momentum complexes for gravity out of the ${\cal P}[\xi]$ currents. For this task the most appropriate Lagrangian is the M\o ller one (\ref{eq:LM}) because it is quadratic in $\ed e^I$. Since the latter is already in the scalar representation, we get
\bea 
{\cal P}_I & = & \frac{\pa L_{\rm M}}{\pa e^I} \os{(\ref{eq:inprodsym})(\ref{eq:starvar})}{=} -\ed e_J \we \star\, F^J_{\,\,\,I}  \\
 & & + \frac{1}{2}\, F^J_{\,\,\,K} \we i_I \star F^K_{\,\,\,J} + \frac{1}{2}\, i_I F^J_{\,\,\,K} \we \star\, F^K_{\,\,\,J}  \nn \\
 & & + \frac{1}{2} \[ \ed e_I \we \star\, F - \frac{1}{2}\, F \we i_I \star F - \frac{1}{2}\,  i_I F \we \star\, F \] \, . \nn  
\eea
the corresponding superpotential is
\beq
{\cal U}_I  \equiv  -\frac{\pa L_{\rm M}}{\pa \ed e^I} \os{(\ref{eq:inprodsym})}{=} e_J \we \star \, F_I^{\,\,\,J} - \frac{1}{2}\, e_I \we \star \, F   \nn \, ,
\eeq
and the equation of motion reads ${\cal P}_I = \ed {\cal U}_I$. An interesting property of this current is that it is traceless in four dimensions
\beq \label{eq:notrace}
e^I \we {\cal P}_I \os{(\ref{eq:partdecomp})}{=} (D-4) \, L_{\rm M} \, ,
\eeq
in complete analogy with YM theory if one uses the GR energy-momentum tensor
\beq \label{eq:notraceYM}
e^I \we T^{\rm YM}_I = (D-4) \, L_{\rm YM} \, .
\eeq
To make contact with known objects, we express ${\cal P}_I$ and ${\cal U}_I$ in terms of the Levi-Civita connection $\hat{\om}_{IJ}$ by using $\ed e^I = -\hat{\om}^I_{\,\,\,J} \we e^J$ and (\ref{eq:partdecomp}), (\ref{eq:bdbprod}) twice
\bea 
{\cal P}_I & = & \frac{1}{2} \[  \hat{\om}^J_{\,\,\,K} \we \hat{\om}^L_{\,\,\,I} \we \ti{e}_{J\,\,\,L}^{\,\,\,\,K} - \hat{\om}^K_{\,\,\,J} \we \hat{\om}^J_{\,\,\,L} \we \ti{e}_{K\,\,\,I}^{\,\,\,\,L} \] \, , \nn \\
{\cal U}_I & = & -\frac{1}{2}\, \hat{\om}^{JK} \we \ti{e}_{JKI}   \, .
\eea
These are known as ``Sparling's $d$-form" and the ``Nester-Witten $(d-1)$-form", respectively \cite{DVM, Sza}. Actually, these names originally refer to forms on the frame bundle and here we have their pullback along a section that is an orthonormal frame. They are tensors under diffeomorphisms but transform inhomogeneously under LLTs and thus provide us with a coordinate-independent definition of energy and momentum $E \equiv \int_{\Si} {\cal P}^0$ and $P^i \equiv \int_{\Si} {\cal P}^i$, for a space-like $\Si \in \De^d$. An important property of this energy-momentum complex, first considered in \cite{DVM}, is the fact that for small enough $\Si$ we have the desired property $E \geq |P^i| \geq 0$, thanks to a direct relation of ${\cal P}_I$ to the Bel-Robinson tensor\footnote{This is shown in \cite{NesPos}, but the authors erroneously call this complex the ``M\o ller complex", since this is not the one proposed by M\o ller in '61, which is a pseudo-tensor under diffeomorphisms.}. To get the general current ${\cal P}[\xi]$, we proceed as in (\ref{eq:Pdecomp})
\beq 
{\cal P}[\xi] = \xi^I {\cal P}_I + \ed\xi^I \we {\cal U}_I \, .
\eeq
So the general superpotential is ${\cal U}[\xi] \equiv \xi^I {\cal U}_I$, i.e. ${\cal P}[\xi] = \ed {\cal U}[\xi]$ when on-shell. The complex considered above corresponds to the choice $\xi = \ep_I$. Another famous energy-momentum complex, the one of M\o ller \cite{Mol}, corresponds to taking a holonomic $\xi$. Let us therefore define some local coordinates $x^{\mu}$, so that the vielbein decomposes $e^I = e^I_{\mu} \ed x^{\mu}$, $\ep_I = e_I^{\mu} \pa_{\mu}$. Then, considering the case $\xi = \pa_{\mu}$ translates into $\xi^I = e^I_{\mu}$ and
\beq \label{eq:EMTMolrel}
{\cal P}_{\mu} \equiv {\cal P}[\pa_{\mu}] = e^I_{\mu} {\cal P}_I + \ed e_{\mu}^I \we {\cal U}_I \, .
\eeq
The second term shows that, although ${\cal U}[\pa_{\mu}]$ is a tensor under diffeomorphisms, this is not the case for ${\cal P}_{\mu}$. To make contact with the usual notation in the literature we use the dual contravariant density representation of the current/superpotential couple
\beq
j^{\mu} \equiv -\frac{1}{2}\, e g^{\mu\nu} \( \star {\cal J}\)_{\nu} \, , \hspace{0.5cm} U^{\mu\nu} \equiv -e g^{\mu\ro} g^{\nu\si}  \( \star\, {\cal U}\)_{\ro\si} \, ,
\eeq
where $e \equiv \det(e_{\mu}^I)$. In terms of these the relation becomes simply
\beq
\left. j^{\mu} \right|_{{\rm EL} = 0}[\xi] = -\frac{1}{2}\, e g^{\mu\nu} \( \star\, \ed {\cal U}[\xi]\)_{\nu} = \na_{\ro} U^{\mu\ro}[\xi] = \pa_{\ro} U^{\mu\ro}[\xi]    \, ,
\eeq
and the conservation equation also makes use of partial derivatives $\pa_{\mu} j^{\mu} = 0$. Note that here $\na$ is the Levi-Civita connection, i.e. the one made out of the Christoffel symbols. This representation is particularly useful in the context of energy-momentum pseudo-tensors where one has a non-trivial dependence on coordinates anyway. In our case, we have
\beq \label{eq:Vden}
U^{\mu\nu}[\xi] \equiv -e g^{\mu\ro} g^{\nu\si}  \( \star\, {\cal U}[\xi]\)_{\ro\si}   \, ,  
\eeq
and for $\xi = \pa_{\mu}$ we obtain M\o ller's superpotential
\beq
U_{\mu}^{\,\,\,\nu\ro} \equiv U^{\nu\ro}[\pa_{\mu}] \os{(\ref{eq:partdecomp})(\ref{eq:bdbprod})(\ref{eq:Hodgesquare})}{=} e \( \de_{\mu}^{\nu} \, \hat{\om}^{\ro} - \de_{\mu}^{\ro} \, \hat{\om}^{\nu} - \hat{\om}_{\mu}^{\,\,\,\nu\ro}  \)  \, ,
\eeq
where
\bea \label{eq:hommunuro}
\hat{\om}_{\mu\nu\ro} & \equiv & e_{\nu}^J e_{\ro}^K \hat{\om}_{\mu JK} = -e_{\nu}^J e_{\ro}^K \( \na_{\mu} e_J^{\ta}\) e_{\ta K} \nn \\
 & = & - e^J_{\nu} \na_{\mu} e_{\ro J} = e^J_{\ro} \na_{\mu} e_{\nu J} \, ,
\eea
$\om_{\mu} \equiv g^{\nu\ro}\om_{\nu\ro\mu}$ and we have used the alternative definition of the Levi-Civita spin connection $\na_{\ep_I} \ep_J = - \hat{\om}_{IJ}^{\,\,\,\,\,\,K} \ep_K$. The so-called ``M\o ller complex" is then given by $M_{\mu}^{\,\,\,\nu} \equiv \pa_{\ro} U_{\mu}^{\,\,\,\nu\ro}$ but now this is not equal to $\na_{\ro} U_{\mu}^{\,\,\,\nu\ro}$ because of the extra $\mu$ index. We therefore retrieve in this representation as well the fact that $M_{\mu}^{\,\,\,\nu}$ is a pseudo-tensor density. However, since the superpotential is a tensor, the definition of energy in the M\o ller case $E \equiv -\int_{\Si} {\cal P}_t = - \int_{\pa \Si} {\cal U}[\pa_t]$, where $g_{tt} \equiv g(\pa_t, \pa_t) < 0$, is invariant under spatial diffeomorphisms.

\subsection{Relation to the Hamiltonian energy}

The notion of energy is not only present in the diffeomorphism Noether charges, but also in the Hamiltonian formalism, and in many cases the two definitions agree. Here we show their relation and its limitations. So let us foliate $\cal M$ with a time coordinate $t$ and use (\ref{eq:diffcurr1}) to write the action in canonical form \cite{NesCovHam} 
\bea
S & \equiv & \int_{\cal M} L = \int_{\cal M} \ed t \we i_{\pa_t} L \\
 & = & \int_{\cal M} \ed t \we \[ \Lie_{\pa_t} \ph^a  \we \frac{\pa L}{\pa \ed \ph^a} + {\cal P}[\pa_t] \] \nn \\
 & \equiv & \int_{\cal M} \[ \dot{\ph}^a  \we \pi_a - \ed t \we {\cal H} \] \, , 
\eea
where we have identified the conjugate momenta $(D-p)$-forms and the Hamiltonian $d$-form
\beq
\pi_a \equiv (-1)^d \frac{\pa L}{\pa \ed \ph^a} \we \ed t \, , \hspace{1cm} {\cal H} \equiv - {\cal P}[\pa_t] \, .
\eeq
Since $\cal H$ is an exact form on-shell, we retrieve what is known from the ADM formalism, namely, that the bulk part of the Hamiltonian is zero on-shell, and the boundary term is thus the superpotential integrated over $\pa \Si_t$. Improving ${\cal P}[\pa_t]$ by adding a total derivative on-shell then amounts to changing that boundary term. In their seminal paper \cite{NesHamBound}, the authors used this relation with the canonical formalism to show that a large class of superpotential complexes actually originates in such Hamiltonian boundary terms, these in turn being determined by the boundary conditions one wishes to impose. This actually alleviated the discomfort in considering pseudo-tensors, because their non-covariant behaviour under diffeomorphisms was ultimately related to the breaking of the symmetry by the choice of boundary conditions.

In our context, it is important to note that this applies to complexes that are defined using holonomic indices, i.e. energy corresponds to a $\mu = t$ component, such as in the case of the M\o ller complex. More precisely, the $t$ parameter which is singled-out in going to the Hamiltonian formalism is the same as the parameter of the symmetry corresponding to ${\cal P}_t$. Of course, for arbitrary time-directions $\xi$ one can always choose the foliation such that $\xi = \pa_t$. But if $\xi$ is field dependent $\xi = \xi[\ph]$, then the canonical term $``\dot{\ph} \we \pi"$ is polluted by $\ph$ and we are out of the canonical formalism. For instance, if $\xi = \ep_0$
\beq
S = \int_{\cal M} e^0 \we i_0 L = \int_{\cal M} e^0 \we \( \Lie_0 \ph^a \we \frac{\pa L}{\pa \ed \ph^a}[\pi] - {\cal H}' \) \, , 
\eeq
where ${\cal H}' \equiv - {\cal P}[\ep_0] \equiv - {\cal P}_0$, then this is not an action in canonical form since the equations of motion of $\ph$ and $\pi$ have non-trivial $\pa e^I$ dependencies. Alternatively, ${\cal H}'$ is a tensor under the full diffeomorphism group. 

So, as could be expected, the tensor ${\cal P}_I$ and the associated charges are intrinsically Lagrangian quantities since they refuse to make the minimal compromise of the coordinate-dependence the canonical formalism requires. We therefore conclude that the Noether currents allow for more general energy definitions than the Hamiltonian formalism since one has also access to field-dependent time-translation generators.

\section{Summary}  \label{sec:conclusion}

In this paper we have shown that for generic field manipulations, including differentiation with respect to fields, one gains in simplicity and efficiency by using exterior calculus. As a first application, we have used this formalism to show the full generality of Noether's theorem, both in terms of currents and charges. We have also used it to show the utility of the ``anholonomic representation" in which vielbein gravity is seen to share many formal properties with standard gauge theories. In particular, we have shown that the partial derivative of the Lagrangian with respect to the vielbein yields a class of Noether currents, just as is the case of Yang-Mills theory. We have shown how some diffeomorphism Noether currents give known energy-momentum complexes of the vielbein. Finally, we have discussed the relation between the Noether energy and other definitions of that notion that are the matter energy-momentum tensor in GR and Hamiltonian boundary terms in the canonical formalism.

\vspace{0.7cm}

\acknowledgements

I would like to thank Michele Maggiore, Stefano Foffa, Yves Dirian and Xiao Xiao for useful discussions. This work is supported by the Swiss National Science Foundation.

\appendix

\section{The M\o ller Lagrangian with forms} \label{sec:GR2}

Here we show the relation between the Einstein-Hilbert Lagrangian 
\beq  \label{eq:LCOmdef}
L_{\rm EH} = \frac{1}{2} \, \hat{\Om}_{IJ}[e] \we \ti{e}^{IJ} \, , \hspace{0.5cm} \hat{\Om}_{IJ} \equiv \ed \hat{\om}_{IJ} + \hat{\om}_I^{\,\,\,K} \we \hat{\om}_{KJ} \, ,
\eeq
and the M\o ller Lagrangian in the form (\ref{eq:LM}). We first integrate by parts 
\bea \label{eq:Lagom1}
L_{\rm EH} & = & \frac{1}{2} \[ \ed \hat{\om}_{IJ} \we \ti{e}^{IJ} + \hat{\om}_I^{\,\,\,K} \we \hat{\om}_{K J} \we \ti{e}^{IJ} \]  \nn \\
 & = & L_{\rm M} + \ed \( \hat{\om}_{IJ} \we \ti{e}^{IJ} \) \, .
\eea
where
\beq
L_{\rm M} \equiv \frac{1}{2} \[ \hat{\om}_{IJ} \we \ed \ti{e}^{IJ} + \hat{\om}_I^{\,\,\,K} \we \hat{\om}_{K J} \we \ti{e}^{IJ} \] \, .
\eeq
The first thing to notice is that the two terms are proportional to each other. Indeed, by definition of the Levi-Civita spin connection, we have $\ed e^I = - \hat{\om}^I_{\,\,\,J} \we e^J$ so
\bea
\hat{\om}_{IJ} \we \ed \ti{e}^{IJ} & \os{(\ref{eq:bexder})}{=} & \hat{\om}_{IJ} \we \ti{e}^{IJK} \we \ed e_K \nn \\
 & = & -\hat{\om}_{IJ} \we \hat{\om}_{KL} \we e^L \we \ti{e}^{IJK} \nn \\
 & \os{(\ref{eq:bdbprod})}{=} & -2\, \hat{\om}_I^{\,\,\,K} \we \hat{\om}_{K J} \we \ti{e}^{IJ}  \, , \nn
\eea
and therefore $L_{\rm M} = -\frac{1}{2} \,\hat{\om}_I^{\,\,\,K} \we \hat{\om}_{K J} \we \ti{e}^{IJ}$. We then express this in terms of the components in the vielbein basis $\om_{IJK} \equiv i_I \om_{JK}$ and $\om_I \equiv \om^J_{\,\,\,JI}$
\bea
L_{\rm M}  & = & -\frac{1}{2} \,\hat{\om}_{AI}^{\,\,\,\,\,\,K} \, \hat{\om}_{BKJ} \, e^{AB} \we \ti{e}^{IJ} \nn \\
 & \os{(\ref{eq:bdbprod})}{=} & -\frac{1}{2} \,\hat{\om}_{AI}^{\,\,\,\,\,\,K} \, \hat{\om}_{BKJ}  \( \et^{AI} \et^{BJ} - \et^{AJ} \et^{BI} \) \ti{e} \nn \\
 & = & \frac{1}{2} \,\( \hat{\om}_{IJK} \, \hat{\om}^{JKI} + \hat{\om}_I \, \hat{\om}^I \) \ti{e}  \, ,
\eea
which, given (\ref{eq:hommunuro}), is the original form of M\o ller's Lagrangian \cite{Mol}. Now, using again $\ed e^I = - \hat{\om}^I_{\,\,\,J} \we e^J$ and (\ref{eq:Hodgedualv}) and (\ref{eq:bdbprod}), we get
\bea
e_I \we \ed e_J \we \star \( e^J \we \ed e^I \) & = &  \hat{\om}_{AJK} \, \hat{\om}^{BIL}  \, e_{AIK} \we \ti{e}^{BJL} \nn \\
 & = &  \( \hat{\om}_{IJK}\, \hat{\om}^{IJK} + \hat{\om}_{IJK} \, \hat{\om}^{JKI} \right. \nn \\
 & & \left. - \hat{\om}_I \, \hat{\om}^I \) \ti{e} \, ,  \nn \\
e_I \we \ed e^I \we \star\( e_J \we \ed e^J \) & = & \hat{\om}_{AIJ} \hat{\om}^{BKL} \, e_{AIJ} \we \ti{e}^{BKL} \nn \\
 & = & 2\, \( \hat{\om}_{IJK} \, \hat{\om}^{IJK} \right. \nn \\
 & & \left. + 2\, \hat{\om}_{IJK} \, \hat{\om}^{JKI} \) \ti{e}  \, , 
\eea
so, defining the 3-forms $F^{IJ} \equiv e^I \we \ed e^J$ and $F \equiv F^I_{\,\,\,I}$, one can write it as (\ref{eq:LM}).

\section{Equivalence of the holonomic and anholonomic representations} \label{sec:stdsveq}

Here we use $\ph^a$ to denote collectively all fields but the vielbein. Using
\bea
\ph^a & = & \frac{1}{p!} \, \ph^a_{I_1 \dots I_p} \, e^{I_1 \dots I_p} \, , \\
\ed \ph^a & = & \frac{1}{p!} \[ \ed \ph^a_{I_1 \dots I_p} \we e^{I_1 \dots I_p} + \ph^a_{I_1 \dots I_p} \, \ed e^{I_1 \dots I_p} \] \, , \nn
\eea
to translate from one representation to the other we get
\bea
\frac{\pa L^{\rm an}}{\pa \ph_{I_1 \dots I_p}^a} & = & \frac{\pa \ph^b}{\pa \ph_{I_1 \dots I_p}^a} \we \frac{\pa L^{\rm hol}}{\pa \ph^b} + \frac{\pa \ed \ph^b}{\pa \ph_{I_1 \dots I_p}^a} \we \frac{\pa L^{\rm hol}}{\pa \ed \ph^b} \nn \\
 & = & \frac{1}{p!} \[ e^{I_1 \dots I_p} \we \frac{\pa L^{\rm hol}}{\pa \ph^a} + \ed e^{I_1 \dots I_p} \we \frac{\pa L^{\rm hol}}{\pa \ed \ph^a} \] \, , \nn \\
\frac{\pa L^{\rm an}}{\pa \ed \ph_{I_1 \dots I_p}^a} & = & \frac{\pa \ph^b}{\pa \ed \ph_{I_1 \dots I_p}^a} \we \frac{\pa L^{\rm hol}}{\pa \ph^b} + \frac{\pa \ed \ph^b}{\pa \ed \ph_{I_1 \dots I_p}^a} \we \frac{\pa L^{\rm hol}}{\pa \ed \ph^b} \nn \\
 & = & \frac{1}{p!} \, e^{I_1 \dots I_p} \we \frac{\pa L^{\rm hol}}{\pa \ed \ph^a} \, ,  \label{eq:piIofpi}
\eea
so the equations of motion for $\ph_{I_1 \dots I_p}^a$ give
\bea
{\rm EL}^{{\rm an}, I_1 \dots I_p}_a & \equiv & \frac{\pa L^{\rm an}}{\pa \ph_{I_1 \dots I_p}^a} - \ed \frac{\pa L^{\rm an}}{\pa \ed \ph_{I_1 \dots I_p}^a} \nn \\ 
 & = & \frac{1}{p!} \[ e^{I_1 \dots I_p} \we \frac{\pa L^{\rm hol}}{\pa \ph^a} + \ed e^{I_1 \dots I_p} \we \frac{\pa L^{\rm hol}}{\pa \ed \ph^a} \right. \nn \\
 & & \left. - \ed \( e^{I_1 \dots I_p} \we \frac{\pa L^{\rm hol}}{\pa \ed \ph^a} \) \] \nn \\
 & = & \frac{1}{p!} \, e^{I_1 \dots I_p} \we \( \frac{\pa L^{\rm hol}}{\pa \ph^a} - (-1)^p \ed \frac{\pa L^{\rm hol}}{\pa \ed \ph^a} \) \nn \\
 & \equiv & \frac{1}{p!} \, e^{I_1 \dots I_p} \we {\rm EL}^{\rm hol}_a \, .
\eea
Thus ${\rm EL}_a = 0$ implies ${\rm EL}^{I_1 \dots I_p}_a = 0$. Since this holds for all $I_1 \dots I_p$ and the vielbein is a basis, we also have that the converse is true. With the same type of manipulations for the vielbein we get
\bea
{\rm EL}_I^{\rm an} - {\rm EL}_I^{\rm hol} & = & \frac{\pa \ph^a}{\pa e^I} \we \frac{\pa L^{\rm hol}}{\pa \ph^a} + \frac{\pa \ed \ph^a}{\pa e^I} \we \frac{\pa L^{\rm hol}}{\pa \ed \ph^a} \nn \\
 & & + \ed \( \frac{\pa \ed \ph^a}{\pa \ed e^I} \we \frac{\pa L^{\rm hol}}{\pa \ed \ph^a} \)  \nn \\
 & = & i_I \ph^a \we \( \frac{\pa L^{\rm hol}}{\pa \ph^a} - (-1)^p \ed \frac{\pa L^{\rm hol}}{\pa \ed \ph^a} \) \nn \\
 & \equiv & i_I \ph^a \we {\rm EL}^{\rm hol}_a  \label{eq:ELdif}
\eea
so they are the same when the rest of the fields are on-shell. We have thus shown that both representations are classically equivalent. Most importantly for us however, the equivalence in the matter sector is independent of the choice of representation for the vielbein. Thus, one can take the equations of motion in the holonomic representation for $\ph^a$ and in the anholonomic one for $e^I$ while still describing the same classical physics. 

Finally, we also show that the Noether currents are the same even though some fields have changed representation with respect to LLTs and diffeomorphisms. We have that the infinitesimal variations decompose
\bea 
\de \ph^a & = & \frac{1}{p!} \, \de \( \ph^a_{I_1 \dots I_p} \we e^{I_1 \dots I_p} \) \nn \\
 & = & \frac{1}{p!}  \( \de \ph^a_{I_1 \dots I_p} \we e^{I_1 \dots I_p} + \ph^a_{I_1 \dots I_p} \we \de  e^{I_1 \dots I_p} \) \nn \\
 & = & \frac{1}{p!} \, \de \ph^a_{I_1 \dots I_p} \we e^{I_1 \dots I_p} \nn \\
 & & + \frac{1}{(p-1)!}\, \ph^a_{I_1 \dots I_p} \we \de e^{I_1} \we e^{I_2 \dots I_p}   \label{eq:varphsv}
\eea
so that the Noether currents of the anholonomic representation read
\bea
{\cal J}^{\rm an} & \equiv &  \de e^I \we \frac{\pa L^{\rm an}}{\pa \ed e^I} + \de \ph^a_{I_1 \dots I_p} \we \frac{\pa L^{\rm an}}{\pa \ed \ph^a_{I_1 \dots I_p}} + K \nn \\
 & \os{(\ref{eq:piIofpi})}{=} &  \de e^I \we \frac{\pa \ed e^J}{\pa \ed e^I}  \we \frac{\pa L^{\rm hol}}{\pa \ed e^J} + \de e^I \we \frac{\pa \ed \ph^a}{\pa \ed e^I}  \we \frac{\pa L^{\rm hol}}{\pa \ed \ph^a} \nn \\
 & & + \frac{1}{p!} \, \de \ph^a_{I_1 \dots I_p} \, e^{I_1 \dots I_p} \we \frac{\pa L^{\rm hol}}{\pa \ed \ph^a} + K  \\
 & = & \de e^I \we \frac{\pa L^{\rm hol}}{\pa \ed e^I} + \frac{1}{p!} \, \de \ph^a_{I_1 \dots I_p} \, e^{I_1 \dots I_p} \we \frac{\pa L^{\rm hol}}{\pa \ed \ph^a}  \nn \\
 & & + \frac{1}{(p-1)!}\, \ph^a_{I I_2 \dots I_p} \de e^I \we e^{I_2 \dots I_p}  \we \frac{\pa L^{\rm hol}}{\pa \ed \ph^a}  + K \nn \\
 & \os{(\ref{eq:varphsv})}{=} & \de e^I \we \frac{\pa L^{\rm hol}}{\pa \ed e^I} + \de \ph^a  \we \frac{\pa L^{\rm hol}}{\pa \ed \ph^a} + K \equiv {\cal J}^{\rm hol} \, .  \nn
\eea

\section{Useful identities} \label{sec:useide}

\begin{enumerate}

%\item
%Levi-Civita symbol contraction
%\beq \label{eq:LeviCivitaProd}
%\vep_{I_1 \dots I_D} \vep^{J_1 \dots J_k I_{k+1} \dots I_D} = - k! (D-k)! \, \de_{[I_1}^{J_1} \dots \de_{I_k]}^{J_k} \, .
%\eeq

\item
Vielbein contraction
\bea 
i_{I_k \dots I_1} e^{J_1 \dots J_l} & = & \frac{l!}{(l-k)!} \, \de_{I_1}^{[J_1} \dots \de_{I_k}^{J_k} e^{J_{k+1} \dots J_l]}  \, , \nn \\
i_{I_1 \dots I_k} \ti{e}_{J_1 \dots J_l} & = & \ti{e}_{J_l \dots J_1I_k \dots I_1} \, . \label{eq:binprod}
\eea

\item
Vielbein exterior derivative
\bea 
\ed e^{I_1 \dots I_k} & = & k \, \ed e^{[I_1} \we e^{I_2 \dots I_k]}  \, , \nn \\
\ed \ti{e}_{I_1 \dots I_k} & = & \ti{e}_{I_1 \dots I_k I_{k+1}} \we \ed e^{I_{k+1}} \, .  \label{eq:bexder}
\eea

\item
$k$-dimensional partial decomposition of $\al \in \Om^p$ with $k \leq p$
\beq \label{eq:partdecomp}
e^{I_1 \dots I_k} \we i_{I_k \dots I_1} \al = \frac{p!}{(p-k)!} \al \, .
\eeq

%\item
%Generalized vielbein matrix inversion
%\beq \label{eq:geninv}
%\vep_{I_1 \dots I_D} e^{I_1}_{\mu_1} \dots e^{I_p}_{\mu_p} = e\, \vep_{\mu_1 \dots \mu_D} e_{I_{p+1}}^{\mu_{p+1}} \dots e_{I_D}^{\mu_D} 
%\eeq

\item
Hodge dual and dual vielbein
\beq \label{eq:Hodgedualv}
\ti{e}^{I_1 \dots I_k} = \star\, e^{I_1 \dots I_k} \, .
\eeq

%\item
%Dual vielbein inversion
%\beq \label{eq:bdinv}
%e^{I_1 \dots I_k} = -\frac{(-1)^{k(D-k)}}{(D-k)!}\, \vep^{I_1 \dots I_D} \ti{e}_{I_{k+1} \dots I_D}
%\eeq

\item
Vielbein - dual vielbein product
\bea 
e^{I_1 \dots I_k} \we \ti{e}_{J_1 \dots J_l} & = & (-1)^{k(l-k)} \frac{l!}{(l-k)!} \nn \\
 & & \times \de_{[J_1}^{I_1} \dots \de_{J_k}^{I_k} \, \ti{e}_{J_{k+1} \dots J_l]} \, . \label{eq:bdbprod}
\eea

\item
Square of the Hodge dual. Let $\al \in \Om^p$, then
\beq \label{eq:Hodgesquare}
\star^2 \al = -(-1)^{p(D-p)} \al \, .
\eeq

\item
Integration by parts of $i_{\xi}$. For all $\al \in \Om^p$ and $\be \in \Om^{D-p+1}$,
\beq \label{eq:intbpi}
i_{\xi} \al \we \be = (-1)^p \, \al \we i_{\xi} \be \, .
\eeq

\item
The product $\al \we \star\, \be$ is symmetric if $\al, \be \in \Om^p$
\beq \label{eq:inprodsym}
\al \we \star\, \be = \be \we \star\, \al \, .
\eeq

%\item
%For $\al \in \Om^p$ and $\be \in \Om^{D-p}$
%\beq
%\star\, \al \we \star\, \be = - \al \we \be
%\eeq

\item
The interior product $i_{\xi}$ is the dual operation of the wedge multiplication by $\xi^{\flat} \equiv g\( \xi, \cdot \) \in \Om^1$
\beq \label{eq:idual}
i_{\xi} \star \al = \star \( \al \we \xi^{\flat} \) \, , \hspace{0.3cm}  \star\, i_{\xi} \al = (-1)^d\, \star \al \we \xi^{\flat} \, .
\eeq

\item
The previous identity implies the following generalizations
\bea 
i_{I_k \dots I_1} \star \al & = & \star \( \al \we e_{I_1 \dots I_k} \)  \, , \nn \\
\star\, i_{I_k \dots I_1} \al & = & (-1)^{kd}\( \star\, \al \) \we e_{I_1 \dots I_k} \, ,  \label{eq:iduale}
\eea

\item
On the other hand
\bea 
\star \( \al \we \ti{e}_{I_1 \dots I_k} \) & = & - (-1)^{(k-p)(D-k)} {k \choose p} \nn \\
 & & \times  e_{[I_{p+1} \dots I_k} \, i_{I_p \dots I_1]}\al    \, . \label{eq:idualtie}
\eea

\item
For all $\al \in \Om^p$ 
\beq \label{eq:deIi}
\( \de e^I \we  \frac{\pa}{\pa e^I} \, i_{I_k \dots I_1} \) \al   = - k\, i_{[I_k} \de e^I \, i_{I |I_{k-1} \dots I_1]} \al \, .
\eeq
%{\it Proof.}
%
%Here we work with the components in some local coordinate system. Since the final result will not depend on $x^{\mu}$, it will also hold globally. We have
%\bea
%& & \( \de e^I \we  \frac{\pa}{\pa e^I} \, i_{I_k \dots I_1} \) \al  \nn\\
% & = & \frac{1}{(p-k)!} \de e^I_{\mu} \( \frac{\pa}{\pa e^I_{\mu}} \, e_{I_k}^{\mu_k} \dots e_{I_1}^{\mu_1} \) \nn \\
% & & \times \al_{\mu_1 \dots \mu_p} \, \ed x^{\mu_{k+1}} \we \dots \we \ed x^{\mu_p} \nn \\
% & = & -\frac{1}{(p-k)!} \de e^I_{\mu} \, e_J^{\mu} e_I^{\nu} \( \frac{\pa}{\pa e^{\nu}_J} \, e_{I_k}^{\mu_k} \dots e_{I_1}^{\mu_1} \) \nn \\
% & & \times \al_{\mu_1 \dots \mu_p} \, \ed x^{\mu_{k+1}} \we \dots \we \ed x^{\mu_p} \nn \\ 
% & = & -\frac{k}{(p-k)!} \de e^I_{\mu} \, e_J^{\mu} e_I^{\nu} \, \de_{\nu}^{\mu_k} \de_{[I_k}^J \, e_{I_{k-1}}^{\mu_{k-1}} \dots e_{I_1]}^{\mu_1}  \nn \\
% & &\times\al_{\mu_1 \dots \mu_p} \, \ed x^{\mu_{k+1}} \we \dots \we \ed x^{\mu_p} \nn \\
% & = & -\frac{k}{(p-k)!} \de e^I_{\mu} \, e_{[I_k|}^{\mu} e_I^{\mu_k} \, e_{I_{k-1}}^{\mu_{k-1}} \dots e_{I_1]}^{\mu_1} \nn \\
% & & \times \al_{\mu_1 \dots \mu_p} \, \ed x^{\mu_{k+1}} \we \dots \we \ed x^{\mu_p} \nn \\
% & = & - k\, i_{[I_k} \de e^I \, i_{I |I_{k-1} \dots I_1]} \al
%\eea
%$\hfill \Box$

\item
For all $\al, \be \in \Om^p$
\beq \label{eq:starvar}
\al \we \( \frac{\pa}{\pa e^I} \, \star \) \be  = \al \we i_I \star \be - (-1)^p i_I \be \we \star\, \al   \, .
\eeq
{\it Proof}.
\bea
& & \de e^I \we \al \we \( \frac{\pa}{\pa e^I} \, \star \) \be = (-1)^p \al \we \( \de e^I \we \frac{\pa}{\pa e^I} \, \star \) \be  \nn \\
 & \os{(\ref{eq:Hodge})}{=} & \frac{(-1)^p }{p!} \[ \al \we \(\de e^I \we  \frac{\pa}{\pa e^I} \ti{e}^{I_1\dots I_p} \) i_{I_p \dots I_1} \be \right. \nn \\
 & & \left. + \al \we \ti{e}^{I_1\dots I_p} \( \de e^I \we  \frac{\pa}{\pa e^I}  i_{I_p \dots I_1} \) \be \] \nn \\
 & \os{(\ref{eq:deIi})}{=} & \frac{(-1)^p }{p!} \[ \al \we \de e^I \we  i_I\ti{e}^{I_1\dots I_p} \, i_{I_p \dots I_1} \be \right. \nn \\
 & & \left. - p\, \al \we \ti{e}^{I_1\dots I_p} \we i_{I_p} \de e^I \we i_{I I_{p-1} \dots I_1}  \be \] \nn \\ 
 & \os{(\ref{eq:intbpi})}{=} & (-1)^p \[ (-1)^p \, \de e^I \we \al \we i_I \( \star \be \) \right. \nn \\
 & & \left. + (-1)^D \frac{1}{(p-1)!}\, i_{I_p}\al \we \ti{e}^{I_1\dots I_p} \we \de e^I \we i_{I I_{p-1} \dots I_1}  \be \] \nn \\
 & = & \de e^I \we \[  \al \we i_I \star \be \right. \nn \\ 
 & & \left. - (-1)^p \frac{1}{(p-1)!} \, i_{I_p} \( \al \we \ti{e}^{I_1\dots I_p} \) \we i_{I I_{p-1} \dots I_1}  \be \] \nn \, .
\eea
We compute the second term separately
\bea
& & i_{I_p} \( \al \we \ti{e}^{I_1\dots I_p} \) \we i_{I I_{p-1} \dots I_1}  \be \nn \\
 & = & i_{I_p} \( \al \we \ti{e}^{I_p I_1\dots I_{p-1}} \) \we i_{I_{p-1} \dots I_1 I}  \be  \nn \\
 & \os{(\ref{eq:Hodgedualv})}{=} &  i_{I_p}\( \al \we \star\, e^{I_p I_1\dots I_{p-1}} \) \we i_{I_{p-1} \dots I_1 I}  \be  \nn \\
 & \os{(\ref{eq:inprodsym})}{=} &  i_{I_p}\( e^{I_p I_1\dots I_{p-1}} \we \star\, \al \) \we i_{I_{p-1} \dots I_1 I}  \be  \nn \\
 & = &  \( i_{I_p}e^{I_p I_1\dots I_{p-1}} \we \star\, \al \right. \nn \\
 & & + \left. (-1)^p e^{I_p I_1\dots I_{p-1}} \we i_{I_p} \star \al \) \we i_{I_{p-1} \dots I_1 I}  \be  \nn \\
 & \os{(\ref{eq:binprod})}{=} & ( D-p+1 ) e^{I_1\dots I_{p-1}} \we \star\, \al \we i_{I_{p-1} \dots I_1 I}  \be  \nn \\ 
 & &  + (-1)^p \, e^{I_p I_1\dots I_{p-1}} \we i_{I_p} \star \al \we i_{I_{p-1} \dots I_1 I}\be  \nn \\
 & = & (-1)^{(p-1)(D-p)} \[  ( D-p+1 ) \star \al \we e^{I_1\dots I_{p-1}} \we i_{I_{p-1} \dots I_1 I }  \be \right.  \nn \\ 
 & & \left. - e^{I_p} \we i_{I_p}  \star \al \we e^{I_1\dots I_{p-1}} \we  i_{I_{p-1} \dots I_1 I }  \be \] \nn \\
 & \os{(\ref{eq:partdecomp})}{=} & (-1)^{(p-1)(D-p)} (p-1)! \[ ( D-p+1 ) \star \al \we i_I \be \right. \nn \\ 
 & & \left. - ( D-p ) \star \al \we  i_I \be \] \nn \\
 & = & (-1)^{(p-1)(D-p)} (p-1)! \star \al \we i_I \be  \nn \\
 & = & (p-1)! \, i_I \be \we \star\, \al \, . \nn
\eea

$\hfill \Box$

\end{enumerate}

\end{document}